\documentclass[article,preprint,groupedaddress]{revtex4}
\usepackage{epsfig}
\usepackage{graphicx}

\begin{document}
\title{Viscosity bound versus the universal relaxation bound}
\author{Shahar Hod}
\affiliation{The Ruppin Academic Center, Emeq Hefer 40250, Israel}
\affiliation{ } \affiliation{The Hadassah Institute, Jerusalem
91010, Israel}
\date{\today}

\begin{abstract}
\ \ \ For gauge theories with an Einstein gravity dual, the AdS/CFT
correspondence predicts a universal value for the ratio of the shear
viscosity to the entropy density, $\eta/s=1/4\pi$. The holographic
calculations have motivated the formulation of the celebrated KSS
conjecture, according to which all fluids conform to the lower bound
$\eta/s \geq 1/4\pi$. The bound on $\eta/s$ may be regarded as a
lower bound on the {\it relaxation} properties of perturbed fluids
and it has been the focus of much recent attention. In particular,
it was argued that for a class of field theories with Gauss-Bonnet
gravity dual, the shear viscosity to entropy density ratio,
$\eta/s$, could violate the conjectured KSS bound. In the present
paper we argue that the proposed violations of the KSS bound are
strongly constrained by Bekenstein's generalized second law (GSL) of
thermodynamics. In particular, it is shown that physical consistency
of the Gauss-Bonnet theory with the GSL requires its coupling
constant to be bounded by $\lambda_{GB}\lesssim 0.063$. We further
argue that the genuine physical bound on the relaxation properties
of physically consistent fluids is $\Im\omega(k>2\pi T)>\pi T$,
where $\omega$ and $k$ are respectively the proper frequency and the
wavenumber of a perturbation mode in the fluid.
\end{abstract}
\bigskip
\maketitle


\section{Introduction}

The anti-de Sitter/conformal field theory (AdS/CFT) correspondence
\cite{Mald,Gub,Witt1,Brig1} has yielded remarkable insights into the
dynamics of strongly coupled gauge theories. According to this
duality, asymptotically AdS background spacetimes with event
horizons are interpreted as thermal states in dual field theories.
This implies that small perturbations of a black hole or a black
brane background correspond to small deviations from thermodynamic
equilibrium in a dual field theory. One robust prediction of the
AdS/CFT duality is a universally small ratio of the shear viscosity
to the entropy density \cite{Poli,Kov1,Buc,Kov2},
\begin{equation}\label{Eq1}
{\eta \over s}={1 \over{4\pi}}\  ,
\end{equation}
for all gauge theories with an Einstein gravity dual in the limit of
large 't Hooft coupling \cite{Noteunit}.

It was suggested in \cite{Kov2} that (\ref{Eq1}) acts as a universal
lower bound [the celebrated Kovtun-Starinets-Son (KSS) bound] on the
ratio of the shear viscosity to the entropy density of general,
possibly nonrelativistic, fluids. Currently this bound is considered
a conjecture well supported for a certain class of field theories
\cite{Notesee,Cher,BekFoux}. So far, all known materials satisfy the
bound for the range of temperatures and pressures examined in the
laboratory. The system coming closest to the bound is the
quark-gluon plasma created at the BNL Relativistic Heavy Ion
Collider (RHIC) \cite{Tean,Adar,Roma1,Song,Notefact}. Other systems
coming close to the bound include superfluid helium and trapped
${^6}$Li at strong coupling \cite{Scha,Rup}. For other related
works, see \cite{SonStar,Mat,Dob,Den,Hern} and references therein.

One naturally wonders: How robust is the conjectured KSS bound? A
remarkable feature of the bound is the fact that it is saturated by
strongly coupled gauge theories with an Einstein gravity dual.
However, it should be emphasized that the Einstein-Hilbert action of
general relativity is the simplest possible gravitational action.
Modifications of the Einstein-Hilbert action are commonly considered
in the modern physical literature, see e.g. \cite{Shap} for a
review. It is well known that quantum corrections of the
gravitational field introduce higher in curvature corrections to the
Einstein-Hilbert action. Such higher-curvature terms naturally arise
in the string theory, see e.g. \cite{Polch,Frolov} and references
therein.

Motivated by the expectations to find higher-curvature corrections
to Einstein gravity in any theory of quantum gravity, Refs.
\cite{Brig1,Brig2} have studied the modification of the ratio
$\eta/s$ due to generic higher-derivative terms in the holographic
gravity dual. For a class of conformal field theories with
Gauss-Bonnet gravity dual, described by an action of the form
\begin{eqnarray}\label{Eq2}
{\cal I}&=&{1 \over {2\ell^3_P}}\int d^5x \sqrt{-g}
\Big[R-2\Lambda\nonumber\\&&
-{3\over\Lambda}\lambda_{GB}(R^2-4R_{\mu\nu}R^{\mu\nu}+
R_{\mu\nu\rho\sigma}R^{\mu\nu\rho\sigma})\Big]\  ,
\end{eqnarray}
it was shown that the shear viscosity to entropy density ratio,
$\eta/s$, could possibly violate the conjectured KSS bound. In
particular, for $(3+1)$-dimensional CFT duals of $(4+1)$-dimensional
Gauss-Bonnet gravity, the viscosity/entropy ratio is given by the
generalized relation \cite{Brig1,Brig2}
\begin{equation}\label{Eq3}
{\eta \over s} = {1 \over{4\pi}}(1-4\lambda_{GB})\  .
\end{equation}

The relation (\ref{Eq3}) implies a possible violation of the KSS
bound (\ref{Eq1}) for positive values of the Gauss-Bonnet coupling
parameter $\lambda_{GB}$. It is important to note that it was later
shown \cite{Brig1} that a self-consist Gauss-Bonnet theory is
characterized by the upper bound
\begin{equation}\label{Eq4}
\lambda_{GB}\leq {9 \over 100}\  .
\end{equation}
The relation (\ref{Eq4}) still permits a large ($36\%$) violation of
the KSS bound (\ref{Eq1}).

One naturally wonders whether there is some subtle inconsistency in
the Gauss-Bonnet theory, a theory which poses a serious challenge to
the validity of the original KSS viscosity bound (\ref{Eq1}).
Revealing an inconsistency in the theory would certainly give
support to the idea of a possible universal lower bound on the ratio
$\eta/s$ for all physically consistent theories. In the present
paper we shall argue that the Bekenstein generalized second law
(GSL) of thermodynamics \cite{Bek1,Haw} may help uncovering such
inconsistencies in the theory. In particular, we shall show below
that the GSL provides an upper bound on the value of the
Gauss-Bonnet coupling parameter $\lambda_{GB}$.

\section{The generalized second law of thermodynamics and the
universal relaxation bound}

It is important to emphasize that hydrodynamics is an effective
theory. In the most common applications of hydrodynamics the
underlying microscopic theory is a kinetic theory. In this case the
microscopic scale which limits the validity of the effective
hydrodynamic description is given by the mean free path
$l_{\text{mfp}}$ \cite{Baier}. More generally, the underlying
microscopic theory is a quantum field theory, which might not
necessarily admit a kinetic description. In these cases, the role of
the parameter $l_{\text{mfp}}$ is played by some typical microscopic
scale like the inverse temperature: $\l_{\text{mfp}}\sim T^{-1}$.
One therefore expects to find a breakdown of the effective
hydrodynamic description at spatial and temporal scales which are of
the order of \cite{Baier}
\begin{equation}\label{Eq5}
l\sim \tau\sim T^{-1}\  .
\end{equation}
Below we shall formulate these order-of-magnitude estimates in a
more accurate way.

One must first identify the physical principle which underlies the
KSS bound (or any other refined bound on the ratio $\eta/s$)
\cite{BekFoux}. The conjectured viscosity/entropy bound is based on
holographic calculations of the shear viscosity for strongly coupled
quantum field theories with gravity duals. These holographic
arguments serve to connect quantum field theory with gravity.
Likewise, the celebrated generalized second law of thermodynamics is
a unique law of physics that bridges quantum theory, gravity, and
thermodynamics, see \cite{Bek1,Haw} for details. The GSL asserts
that in any interaction of a black hole with an ordinary matter, the
sum of the entropies (matter+hole) never decreases \cite{Bek1,Haw}.

One of the most remarkable predictions of the GSL is the existence
of a universal entropy bound \cite{Bek4,Bek5}. Furthermore, the GSL
allows one to derive in a simple way two important new quantum
bounds:
\begin{itemize}
\item{The universal relaxation bound \cite{Hod1,Hod2,Gruz,Hodn1,Hodn2}. This bound asserts that the
relaxation time of a perturbed thermodynamic system is bounded from
below by \cite{Noterela}
\begin{equation}\label{Eq6}
\tau\geq 1/\pi T\  ,
\end{equation}
where $T$ is the temperature of the system. This bound can be
regarded as a quantitative formulation of the third law of
thermodynamics. One can also write this bound as \cite{Noterlx}
\begin{equation}\label{EqNew1}
\Im \omega\leq \pi T\  ,
\end{equation}
where $\omega$ is the frequency of the perturbation mode [see Eq.
(\ref{Eq8}) below].}
\item{A closely related conclusion is that thermodynamics can not be
defined on arbitrarily small length scales. In particular, it was
shown in \cite{Pes1,BekFoux} that the minimal length scale (radius)
$\ell$ for which a consistent thermodynamic description is available
for a fluid with zero chemical potential is given by
\begin{equation}\label{Eq7}
\ell_{\text{min}}=1/2\pi T\  .
\end{equation}
}
\end{itemize}
It is worth noting that the bounds (\ref{Eq6}) and (\ref{Eq7}) are
consistent with the heuristic argument which lead to the approximate
relation (\ref{Eq5}).

\section{An upper bound on the Gauss-Bonnet coupling parameter
$\lambda_{GB}$}

Consider a parcel of fluid with an effective size $\ell$. We shall
now analyze the response of the fluid to small mechanical
perturbations. The longest perturbation mode which can fit into a
space region of effective radius $\ell$ is characterized by a
wavelength $\lambda_{\text{max}}=2\pi\ell$. Taking cognizance of the
spatial bound (\ref{Eq7}) \cite{Pes1,BekFoux}, this perturbation
mode is characterized by the minimal wavenumber
\begin{equation}\label{EqNew2}
k_{\text{min}}=2\pi T\  .
\end{equation}

The response of a medium to mechanical excitations is characterized
by two types of normal modes, corresponding to whether the momentum
density fluctuations are transverse or longitudinal to the fluid
flow. Transverse fluctuations lead to the shear mode, whereas
longitudinal momentum fluctuations lead to the sound mode
\cite{Notethere}. These perturbation modes are characterized by
distinct dispersion relations which describe the poles positions of
the corresponding retarded Green functions \cite{SonStar}.

It is well known that the viscosity coefficient $\eta$ characterizes
the intrinsic ability of a perturbed fluid to {\it relax} towards
equilibrium [see Eq. (\ref{Eq8}) below]. It is therefore quite
plausible \cite{Hodn1} that there is an underlying physical
connection between a lower bound on viscosity (such as the
conjectured KSS bound or any other refined version of it) and the
thermodynamic lower bound (\ref{Eq6}) on {\it relaxation} times of
physical systems.

Let us examine the behavior of the shear relaxation mode for fluids
with zero chemical potential. The Euler identity reads
$\epsilon+P=Ts$, where $\epsilon$ is the energy density, $P$ is the
pressure, $T$ is the temperature, and $s$ is the entropy density of
the fluid. The dispersion relation for a shear wave with frequency
$\omega$ and wavenumber $k\equiv 2\pi/\lambda$ is given by
\cite{Baier,Nats}:
\begin{equation}\label{Eq8}
\omega(k)_{\text{shear}}=i{{\eta}\over{Ts}}k^2+i\tau_{\text{shear}}\Big({{\eta}\over{Ts}}\Big)^2k^4
+O\Big({{k^6}\over{T^5}}\Big)\  ,
\end{equation}
where $\eta$ is the shear viscosity coefficient of the standard
first-order hydrodynamics, and $\tau_{\text{shear}}$ is a relaxation
coefficient which reflects contributions from second and third order
causal hydrodynamics
\cite{Mul,Isr,IsrSte,Baier,Nats,Kapu,Spring,Heller,Hub,Beni,BuchMy,KovStarn,Notecorr}.

The imaginary nature of the dispersion relation (\ref{Eq8}) entails
a damping ({\it relaxation}) of the perturbation mode. Its magnitude
(which explicitly depends on the viscosity/entropy ratio) therefore
quantifies the intrinsic ability of a fluid to dissipate
perturbations and to approach thermal equilibrium.


The AdS/CFT correspondence provides a powerful tool which allows one
to compute (numerically) the exact dispersion relation
$\omega(k)_{\text{shear}}$ of strongly coupled models. In
particular, we are interested in determining the functional
dependence of the dispersion relation
$\omega(k;\lambda_{GB})_{\text{shear}}$ on the Gauss-Bonnet coupling
$\lambda_{GB}$. This can be achieved with the help of the
guage/gravity correspondence, see \cite{KovStarn,BuchMy} for
details.

The thermodynamic bounds (\ref{Eq6}) and (\ref{Eq7}) are obviously
two faces of the same physical principle (In fact, both these bounds
are based on the GSL \cite{Bek4,Bek5}). In particular, it is
physically expected that a fluid system whose size $\ell$ violates
the spatial bound (\ref{Eq7}) would also violate the temporal
relaxation bound (\ref{Eq6}). Explicitly, the breakdown of the
effective hydrodynamic description for a system with spatial size
$\ell <1/2\pi T$ [see Eq. (\ref{Eq7})] is expected to be reflected
in the shear dispersion relation which characterizes the relaxation
properties of the perturbed system. As explained above, the
perturbation modes of a system which violates the spatial bound
(\ref{Eq7}) are all characterized by wavenumbers $k$ which are {\it
larger} than $2\pi T$ [see Eq. (\ref{EqNew2})]. Thus, a violation of
the temporal relaxation bound (\ref{Eq6}) by a system which violates
the spatial bound (\ref{Eq7}) is expected to be reflected by the
physical inequality [see Eq. (\ref{EqNew1})]
\begin{equation}\label{Eq9}
\Im\varpi({\rm q}>1)>{1 \over 2}\  ,
\end{equation}
where $\varpi\equiv\omega/2\pi T$ and ${\rm q}\equiv k/2\pi T$.

An upper bound on the physically allowed values of the Gauss-Bonnet
coupling parameter $\lambda_{GB}$ can be inferred by substituting
${\rm q}=1$ in the exact shear dispersion relation
$\varpi=\varpi({\rm q};\lambda_{GB})_{\text{shear}}$ (this
dispersion relation is computed numerically, see
\cite{KovStarn,BuchMy} for details) and imposing the inequality
$\Im\varpi>1/2$ [see Eq. (\ref{Eq9})] for this limiting value of the
perturbation wavenumber.

In particular, using the dimensionless relations
$\eta/s=(1-4\lambda_{GB})/4\pi$ [see Eq. (\ref{Eq3})] and
$\tau_{\text{shear}}T(\lambda_{\text{GB}}=0)=(2-\ln2)/2\pi$
\cite{BuchMy}, one finds from figure 1 of \cite{BuchMy} that, in the
physical regime $\lambda_{GB}\in[-0.711(2),0.113(0)]$, the
dimensionless physical parameter
$[\tau_{\text{shear}}T](\lambda_{\text{GB}})$ is well fitted by the
simple functional expression
\begin{equation}\label{Eq10}
[\tau_{\text{shear}}T](\lambda_{\text{GB}})=a\cdot
\lambda^2_{\text{GB}}+b\cdot \lambda_{\text{GB}}+c\ \ \ \
\text{with}\ \ \ \ a=-0.607(8)\ ,\ b=-1.000(1)\ , \
c={{2-\ln2}\over{2\pi}}\  .
\end{equation}
Substituting Eqs. (\ref{Eq3}) and (\ref{Eq10}) into the shear
dispersion relation (\ref{Eq8}) \cite{Notecorr}, and imposing the
characteristic inequality (\ref{Eq9}) [which, as discussed above,
stems from the GSL \cite{Bek1,Haw} and its closely related physical
bounds (\ref{Eq6}) and (\ref{Eq7})] for this limiting value of the
wavenumber, one finds the upper bound
\begin{equation}\label{Eq11}
\lambda_{GB}\lesssim 0.063\
\end{equation}
on the physically allowed values of the Gauss-Bonnet coupling
parameter. We note that this bound is stronger than the original
constraint (\ref{Eq4}) \cite{Notesim,NR}.

\section{Summary}

In summary, in this paper we have given support to the idea that the
Bekenstein generalized second law of thermodynamics (GSL) may
constrain possible violations of the KSS viscosity bound
(\ref{Eq1}). In particular, assuming the validity of the GSL [and
its closely related physical bounds (\ref{Eq6}) and (\ref{Eq7})],
one finds the upper bound (\ref{Eq11}) on the value of the
Gauss-Bonnet coupling parameter $\lambda_{GB}$. This, in turn,
constrains the possible violations of the original KSS bound by CFT
plasmas with Gauss-Bonnet gravity duals to be less than $\sim 25\%$
\cite{Note12}.

Finally, we would like to emphasize that it is quite plausible that
there is a deep physical connection between the original KSS lower
bound on the ratio $\eta/s$ (which characterizes the {\it
relaxation} properties of a perturbed fluid system) and the
universal {\it relaxation} bound (\ref{Eq6}). However, the two
bounds are not identical. In particular, the KSS bound (\ref{Eq1})
is known to be violated in the model problem studied in the present
paper. We would therefore like to {\it conjecture} that the genuine
physical bound on the relaxation properties of physically consistent
fluids with zero chemical potential is given by
\begin{equation}\label{Eq12}
\Im\varpi({\rm q}>1)_{\text{shear}}>{1 \over 2}\  .
\end{equation}
To our best knowledge, the conjectured relaxation bound (\ref{Eq12})
is respected by all known fluids. It is worth noting that the
original KSS bound (\ref{Eq1}) seems to be a sufficient (but not a
necessary) condition \cite{Note111} for the validity of the proposed
relaxation bound (\ref{Eq12}).

\bigskip
\noindent
{\bf ACKNOWLEDGMENTS}
\bigskip

This research is supported by the Carmel Science Foundation. I thank
Yael Oren, Arbel M. Ongo, Ayelet B. Lata, and Alona B. Tea for
stimulating discussions.



\begin{thebibliography}{99}
\bibitem{Noteunit} We use natural units for which $G=c=\hbar=k_B=1$.
\bibitem{Notefact} In fact, it was the challenge presented by the
quark-gluon plasma which motivated the activity leading to the
formulation of the KSS bound (\ref{Eq1}).
\bibitem{Noterela} The physical connection between the universal relaxation bound (\ref{Eq6})
and the Bekenstein entropy bound \cite{Bek4,Bek5} was established in
\cite{Hod1}.
\bibitem{Noterlx} The relaxation time of a perturbed physical system is given by the reciprocal of
the imaginary part of the perturbation frequency,
$\tau=1/\Im\omega$.
\bibitem{Notethere} There is also the diffuse mode in the presence of a conserved current.
\bibitem{Notecorr} We note that the correction term in (\ref{Eq8}) becomes small in the $\eta/s\ll 1$
limit, the case of most interest here. In fact, for $\lambda_{GB}=0$
and $k=2\pi T$, the truncated dispersion relation (\ref{Eq8}) agrees
with the exact (numerical) result to within $3.6\%$ \cite{KovStarn}.
\bibitem{Notesim} It is interesting to note that, in superconformal
gauge theories with non-equal central charges $c\neq a$, there are
known violations of the KSS viscosity bound (see e.g., \cite{NR}).
Intriguingly, the sign of $(c-a)/a$ is positive for these physical
systems. As pointed out by the anonymous referee, it may be
interesting to explore the possible relation between the GSL and the
sign of the dimensionless ratio $(c-a)/a$ in these theories. We
would like to emphasize that in order to investigate the possible
implications of the GSL constraints on the sign of the physical
parameter $(c-a)/a$, one should first determine (probably using
numerical techniques) the resonance dispersion relation
$\omega=\omega(k;a,c)$ of these physical systems. This issue
deserves further investigation and is beyond the scope of the
present paper.
\end{thebibliography}
\end{document}